\documentclass[10pt]{iopart}

\usepackage{cite}
\usepackage[english]{babel}
\usepackage[utf8]{inputenc} 
\usepackage[table]{xcolor} 
\usepackage[%
  colorlinks=true,
  urlcolor=blue,
  linkcolor=blue,
  citecolor=blue
]{hyperref}
\usepackage{algorithm} 
\usepackage{algpseudocode} 
\usepackage{bbold} 
\usepackage{xcolor} 
\usepackage{graphics} 
\usepackage{amssymb}
\usepackage{bm}
\usepackage{bbm}
\usepackage{tabularx, booktabs}
\usepackage{xspace}
\usepackage{listings}
\usepackage{lipsum}
\usepackage{chemformula} 
\usepackage{braket}
\usepackage[lofdepth]{subfig}
 \usepackage{ulem}
\newcolumntype{Y}{>{\centering\arraybackslash}X}
\definecolor{dgreen}{rgb}{0,.5,0}
\definecolor{dblue}{rgb}{0,0,.5}
\definecolor{dred}{rgb}{0.5,0,.5}
\usepackage{soul}
 
\newcommand{\mel}[3]{\langle #1 | #2 | #3 \rangle}

\begin{document}

\title{Combining  Effective Hamiltonians and Brillouin-Wigner Approach: A Perturbative Approach to Spectroscopy}

\author{Oussama Bindech}
\address{Laboratoire de Chimie Quantique, Institut de Chimie,
CNRS/Université de Strasbourg, 4 rue Blaise Pascal, 67000 Strasbourg, France}

 \author{Bastien Valentin} 
\address{Laboratoire de Chimie Quantique, Institut de Chimie,
CNRS/Université de Strasbourg, 4 rue Blaise Pascal, 67000 Strasbourg, France}

\author{Saad Yalouz*}
\ead{syalouz@unistra.fr}
\address{Laboratoire de Chimie Quantique, Institut de Chimie,
CNRS/Université de Strasbourg, 4 rue Blaise Pascal, 67000 Strasbourg, France}

\author{Vincent Robert*} 
\ead{vrobert@unistra.fr}
\address{Laboratoire de Chimie Quantique, Institut de Chimie,
CNRS/Université de Strasbourg, 4 rue Blaise Pascal, 67000 Strasbourg, France}




\begin{abstract}
The numerical cost of variational methods suggests using perturbative approaches to determine the electronic structure of molecular systems.
In this work, a sequential construction of effective Hamiltonians drives the definition
of approximate model functions and energies in 
a multi-state Rayleigh-Schrödinger perturbative 
scheme. A second step takes advantage of 
an updated partitioning of the Hamiltonian
to perform a state-specific Brillouin-Wigner energy correction
based on a well-tempered perturbation expansion. 
The  multi-step RSBW method is exemplified on model-Hamiltonians to stress its robustness, efficiency 
and applicability to spectroscopy determination.

\end{abstract}

 \maketitle
 \ioptwocol

%
%
%

%

%

\section{Introduction} 

A key point in the study of strongly correlated systems is selecting a computational method that can balance computational feasibility with a high-level description of many-electron effects. 
To achieve spectroscopic accuracy, a Full Configuration Interaction (FCI) expansion of the wavefunction is theoretically the most precise option possible.
While many recent works have focused on developing variants and extensions of this approach (see, for example, Refs.~\cite{yao2021orbital, kossoski2022hierarchy, damour2022ground, eriksen2020shape, garniron2019quantum, coe2023analytic, tubman2020modern}), FCI-like methods remain computationally prohibitive due to their exponential numerical cost,
prompting the search for alternative approaches. 
One such alternative is the Complete-Active-Space Self-Consistent Field (CASSCF) approach~\cite{siegbahn1980comparison, siegbahn1981complete, roos1980complete}, which leverages a Complete Active Space (CAS) wavefunction expansion combined with orbital optimization techniques.
In practice, this method is considered a reference tool for capturing most of the so-called ``static correlation'', inspiring numerous recent developments focused on orbital optimization.~\cite{ding2023quantum, yalouz2021state, mizukami2020orbital, mahler2021orbital, ammar2023bi, yalouz2023orthogonally, barca2018simple, wouters2014density, keller2015efficient, liu2013multireference}
However, even within such methods, achieving spectroscopic accuracy remains a key issue due to the missing electronic ``dynamical'' correlation contributions.
 To address this, perturbative treatments offer a systematic order-by-order expansion to introduce 
 the missing contributions. 
 In perturbative theory, the exact Hamiltonian \(\hat{H}\) is divided into a reference \(\hat{H}_0\) and a perturbation \(\hat{W}_0\), such that \(\hat{H} = \hat{H}_0 + \hat{W}_0\). The success of this approach depends on \(\hat{H}_0\)'s ability to produce accurate reference eigenstates and eigenvalues. Practically, \(\hat{H}_0\) must include the dominant contributions while remaining simple enough to be computationaly tractable. 
Challenges, however, may arise due to quasi-degenerate states, which can lead to vanishing energy denominators and divergence in the expansion series. 
Besides,  
the presence of so-called intruder states can also be problematic in a Complete Active Space Second-Order Perturbation Theory
treatment.~\cite{roos1982simple,andersson1990second}
This issue can be elegantly mitigated by using the N-electron Valence State Perturbation Theory (NEVPT2).~\cite{angeli2001introduction,angeli2001n,angeli2002n} 

The leading perturbative methods rely on Brillouin-Wigner (BW) or Rayleigh-Schrödinger (RS) schemes.~\cite{li2023toward,yi2019multireference,hubavc2010brillouin,hubavc1994size,wilson2003brillouin} 
In both approaches, the splitting and related definition of the model (e.g., active space) and orthogonal 
(so-called perturbers) spaces  might not be straightforward.
While the former does not suffer from the intruder states issue, it is much less widely used, possibly because of the energy-dependent structure of the energy expansion and the size-extensivity error.~\cite{carter2023optimizing}
Despite this, the systematic order-by-order expansion to achieve state-specific energy corrections makes it very appealing. 

The Epstein-Nesbet partition, a variant of the RS approach with a zero-diagonal
perturbation,
\cite{nesbet1955configuration,nesbet1955configuration2}
has proven to be limited when states are expressed as linear combinations
of non-degenerate configurations.
A significant improvement was brought by the 
Configuration Interaction using a Perturbative Selection done Iteratively (CIPSI) method. It is to be considered as 
a landmarck in RS expansions when  multiconfigurational wavefunction are 
necessary.\cite{huron1973iterative}

More recently, the combination of RS and BW theories has provided an original two-step method (referred to as RSBW) to progressively include perturbation effects.~\cite{delafosse2024two}
The RSBW procedure follows the reconstruction of the zeroth-order Hamiltonian from the effective Hamiltonian diagonalization, thereby implementing part of the perturbation effects in the subsequent BW expansion. 
Even though the relevance of the method was evaluated, the re-partitioning of the Hamiltonian may not be systematic enough to guarantee an improved convergence in the BW energy expansion. A prerequisite in the RSBW method is a clear-cut definition of the model space to construct and diagonalize an effective Hamiltonian. 

In this work, we propose an extension of the RSBW method using a multi-step approach to identify the configuration subspaces that necessitate RS treatment. The method employs a scanning procedure across the spectrum by constructing successive effective Hamiltonians. This approach enables (\textit{i}) accounting for static correlation effects and (\textit{ii}) progressively mitigating the influence of the updated perturbation on the redefinition of the zeroth-order Hamiltonian.
In the final step, a state-specific BW energy expansion is conducted to deliver a perturbative approach to the spectroscopy of model systems.
This method can be considered as an alternative to demanding full diagonalization, leveraging the individual strengths of the RS and BW approaches.

In a first section of this paper, an overview of the
previously reported RSBW method is given to
stress its limitation and to motivate the
multi-step RSBW extension.
Then, the  method is applied to  parameterized Hamiltonians.

\section{Construction of a well-tempered perturbation expansion.}

\subsection{RSBW method: an overview} 

Recently, a two-step perturbative approach
has been proposed by taking advantage of the
RS and BW schemes.~\cite{delafosse2024two} 
The RSBW method starts with the definition of an appropriate model space $P$
constructed on model configurations
$\lbrace |{\alpha}\rangle \rbrace$, which are eigenfunctions of the unperturbed Hamiltonian $\hat{H}_0$~\cite{lindgren2012atomic}:
\begin{equation}
    \hat{H}_0 | \alpha \rangle = \mathcal{E}_{\alpha} | \alpha \rangle. 
\end{equation}
The diagonalization of the second-order effective Hamiltonian in the $P$-space produces 
model functions 
$\{ | {\Psi^{\mbox{\tiny RS}}_\alpha} \rangle \}$, energies 
$\{ E^{\mbox{\tiny RS}}_\alpha \}$,
and a different partitioning  $\hat{H} = \hat{H}^{\mbox{\tiny RS}} + \hat{W}$  with
\begin{equation}
    \hat{H}^{\mbox{\tiny RS}} = \sum_{\alpha \in P} E^{\mbox{\tiny RS}}_\alpha \ |{\Psi^{\mbox{\tiny RS}}_\alpha}\rangle\langle{\Psi^{\mbox{\tiny RS}}_\alpha}| +   \sum_{\beta \in Q}  E^{\mbox{\tiny RS}}_{\beta} |{\beta}\rangle\langle{\beta}|.
    \label{eq:HRS_def}
\end{equation}
The so-called perturbers $\{|{\beta} \rangle \}$  with energies $E^{\mbox{\tiny RS}}_{\beta} = \mathcal{E}_{\beta} = \mel{\beta}{\hat{H}_0}{\beta}$ span the
orthogonal $Q$-space. 
The impact of $\hat{W} = \hat{H}-\hat{H}^{\mbox{\tiny RS}}$ is  reduced as compared to that of $\hat{W}_0 = \hat{H}-\hat{H}_0$ since part of
the perturbation contributions are included in the effective Hamiltonian: 
this is the essence of the RSBW method.
The missing contributions are then included
in a truncated BW energy expansion where the exact energy 
$E^{\rm exact}_i$
is approximated as 
$E^{\mbox{\tiny RSBW}}_i$ and the energy denominators are set  
to $E^{\mbox{\tiny RSBW}}_i - E^{\mbox{\tiny RS}}_j$ 
in the m$^{\mbox{\tiny th}}$ order
 expansion of the resolvent:
\begin{equation} 
    \hat{\Omega}^{\left(m \right)}_i = 
   \left( \sum_{j \neq i} \frac{|{\Psi_j}\rangle\langle {\Psi_j} |  }{E^{\rm exact}_i - \hat{H}^{\mbox{\tiny RS}}} \hat{W} \right)^{\!\!m}. 
    \label{eq:BW_general2}
\end{equation}
Despite its performance reflected by an improved convergence,
the RSBW method relies on the identification of a convenient model space which remains a bottleneck (intruder state issue).
Thus, we considered that such limitation could be overcome by a systematic screening
of the states energies and couplings, 
to step-by-step build and diagonalize effective Hamiltonians before 
applying a final BW treatment.

\subsection{Multi-step RSBW method}

Starting from the ground state 
energy
$\mathcal{E}_{0}$ of  $\hat{H}_0$,
a screening of the zeroth-order eigenenergies $\lbrace \mathcal{E}_{\gamma}  \rbrace$
is performed through the evaluation
of the ratios
\begin{equation} 
    \rho_{0\gamma,1} = \left\lvert 
    \frac{ \mel{0}{\hat{W}_0}{\gamma} }
    {\mathcal{E}_{0} - \mathcal{E}_{\gamma}}
        \right\rvert,
    \label{eq:RS_criterion_step1}
\end{equation}
where $|{0} \rangle$ and  $|{\gamma} \rangle $ are the corresponding eigenvectors, and the 
extra index refers
to the first RS transformation.  
 The objective is to identify the $|{\gamma} \rangle$'s that fulfill  
$\rho_{0\gamma,1} > \rho_{\rm min}$, with $\rho_{\rm min}$ an arbitrary threshold value.
Such criterion emerging from  perturbation theory 
in quantum mechanics is the guiding one for the progressive sub-space definition
in the CIPSI method.\cite{huron1973iterative}
This procedure makes it possible to develop a model space $P_1$ based on a perturbation criterion. 
Evidently, the smaller the
$\rho_{\rm min}$ value, the larger the dimension $d_1$ of the $P_1$-space,
and strictly degenerate states 
(\textit{i.e.} $\mathcal{E}_{\gamma} = \mathcal{E}_{0}$) 
are automatically retained following the
Bloch theory. 
If the $P_1$-space dimension is unity after this procedure ($d_1 =1$), a similar
screening is conducted from $\mathcal{E}_{1}$, and possibly from higher-lying
eigenstates of $\hat{H}_0$, until a model space with $d_1 \ge 2$  is identified.
The states which do not belong to $P_1$ are referred to as perturbers $\ket{\beta}$.
Where this condition is not met,  one would be tempted either  to
use standard
perturbation theory, or to reduce the threshold value $\rho_{\rm min}$.
In practice, challenging issues stem from quasi-degenerate states in some energy windows (\textit{e.g.} singlet-triplet energy difference
in the low-energy spectrum).
Therefore, the screening effort might be rather limited to concentrate on a selection
of states
with the identification of possible intruders
in the $\hat{H}_0$ spectrum. 
Following the RSBW method, (\textit{i}) a second-order effective Hamiltonian is built
and diagonalized in the identified model space $P_1$, and (\textit{ii}) an updated zeroth-order
Hamiltonian  $\hat{H}^{\mbox{\tiny RS}}_1$  (see Eq.~\ref{eq:HRS_def})
and perturbation $ \hat{W}_1$ are  defined.

  \begin{figure*}[!ht] 
     \centering
     \includegraphics[width=16.5cm]{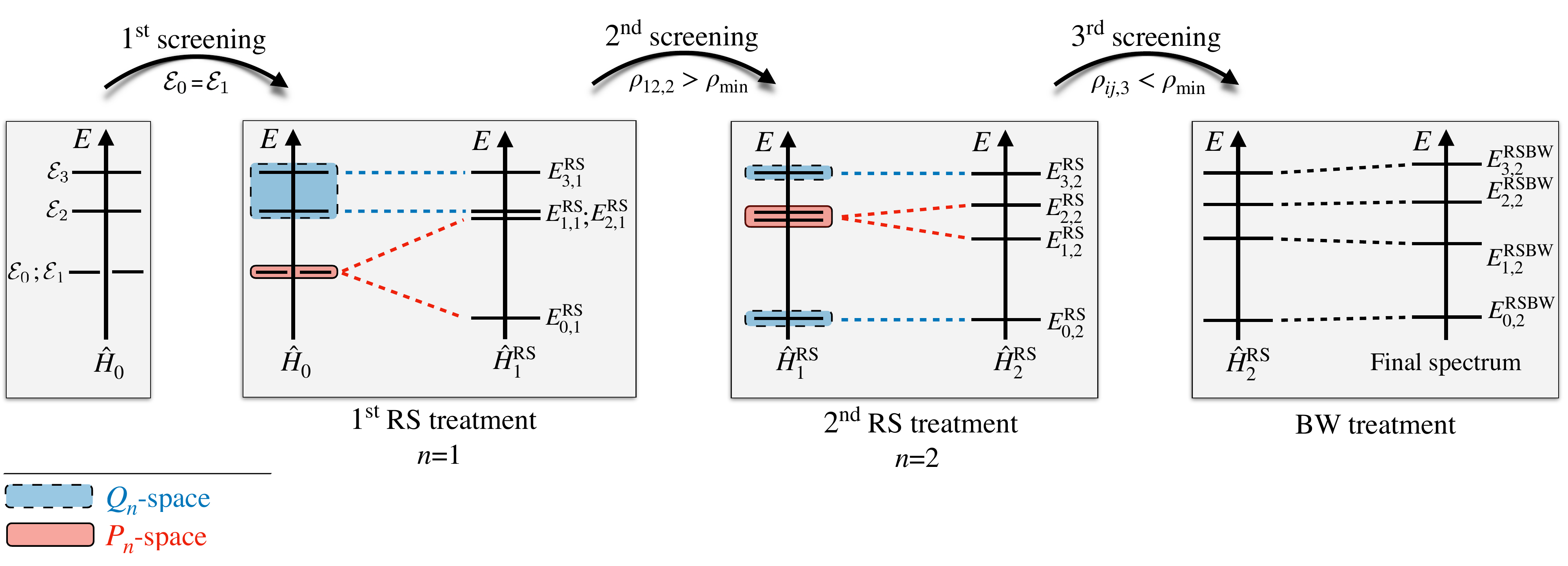}
     \caption{ Description of the multi-step RSBW method. For the sake of simplicity, only two
     steps are shown. 
     The first one uses a $\hat{H} = \hat{H}_0 + \hat{W}_0$ partitioning,
     whereas the second step takes advantage of the redefinition of the zeroth-order Hamiltonian following the RSBW strategy $\hat{H} = \hat{H}^{\mbox{\tiny RS}}_1 + \hat{W}_1$.
     At each step $n$, the screening procedure identifies the  $|{\Psi^{\mbox{\tiny RS}}_{j,n}}\rangle$ states that 
     for a given $|{\Psi^{\mbox{\tiny RS}} _{i,n}}\rangle$ state are either strictly degenerate or fulfill
     $\rho_{ij,n} > \rho_{\rm min}$ to build up the
     $P_n$ and $Q_n$ spaces. 
     In the final step, the eigenstates and eigenenergies of $\hat{H}^{\mbox{\tiny RS}}_2$  provide an improved
     zeroth-order states and zeroth-order energies of the system perturbed by $\hat{W}_2$ used for the final BW treatment.
     } 
     \label{fig:scheme_iterRSBW}
 \end{figure*}


This procedure is then repeated as shown in Figure~\ref{fig:scheme_iterRSBW} (restricted
to two steps for illustration). 
In the following,  we shall 
use  the notations $\{E^{\rm RS}_{i,n} \}$  and $\{E^{\rm RSBW }_{i,n} \}$ to refer to the approximate eigenenergies obtained  from the RS and the RSBW treatments, respectively. 
In this labelling, the index $i$ is used to reference the approximate eigenstates
$\{|\Psi^{\rm RS}_{i,n} \rangle \}$
of the system and $n$ refers to the number of RS steps. 
At step $n+1$, the screening   is performed from the 
 partitioning $\hat{H} = \hat{H}^{\mbox{\tiny RS}}_n + \hat{W}_n$ with the RS Hamiltonian 
 \begin{equation}
 \hat{H}^{\mbox{\tiny RS}}_n = \sum_{\alpha \in \text{$P_n$}} E^{\mbox{\tiny RS}}_{\alpha,n} \ |{\Psi^{\mbox{\tiny RS}}_{\alpha,n}}\rangle\langle{\Psi^{\mbox{\tiny RS}}_{\alpha,n}}| + \sum_{\beta \in \text{$Q_n$}} E^{\mbox{\tiny RS}}_{\beta,n} \ |{\Psi^{\mbox{\tiny RS}}_{\beta,n}}\rangle\langle{\Psi^{\mbox{\tiny RS}}_{\beta,n}}|
     \label{eq:HRS_iteration_n}
 \end{equation}
(with $\hat{H}^{\mbox{\tiny RS}}_0 = \hat{H}_0 $)
to generate the next $P_{n+1}$- and  $Q_{n+1}$-spaces under the
condition:
\begin{equation} 
    \rho_{ij,n+1} = \left\lvert 
    \frac{ \mel{\Psi^{\mbox{\tiny RS}}_{i,n}}{\hat{W}_{n}}{\Psi^{\mbox{\tiny RS}}_{j,n}} }
    {E^{\mbox{\tiny RS}}_{i,n} - E^{\mbox{\tiny RS}}_{j,n}}
        \right\rvert > \rho_{\rm min}.
    \label{eq:RS_criterion_step_n}
\end{equation}
In all notations, the extra lower index $n$ indicates the number of performed RS transformations.
Since the diagonalization is restricted to the $P_{n+1}$ model space,
the orthogonal $Q_{n+1}$ states energies (\textit{i.e.} the perturbers  $|\Psi^{\mbox{\tiny RS}}_{\beta,n} \rangle $ energies) 
remain identical to the ones at step $n$,  
 $E^{\mbox{\tiny RS}}_{\beta,n+1} = E^{\mbox{\tiny RS}}_{\beta,n} = 
 \mel{\Psi^{\mbox{\tiny RS}}_{\beta,n}}{\hat{H}^{\mbox{\tiny RS}}_n}{\Psi^{\mbox{\tiny RS}}_{\beta,n}}
 $.
Two important points should be made regarding the strategy. First, the $\rho_{\rm min}$ value directly controls the size $d_n$ of the 
effective Hamiltonian to be diagonalized in the $P_n$-space.
Then,  the status
of a given state  (model or perturber) may change along the
construction of the successive model spaces (see Figure~\ref{fig:scheme_iterRSBW}).
However, previous inspections~\cite{delafosse2024two}  suggest that the
perturbation contribution $\hat{W}_n$ decreases along this pre-conditioning of the 
Hamiltonian partitioning.
Thus, the dimensions of the screened $P_{n+1}$-spaces should all be unity ($d_{n+1} =1$)
after the construction
and diagonalization of a limited number of 
effective Hamiltonians.
Guided by the RSBW method~\cite{delafosse2024two}, the last step consists in a state-specific BW expansion of the 
energies, with well-tempered perturbation contributions (see right part in Figure \ref{fig:scheme_iterRSBW}).
 
{ To conclude this section and aid the reader's comprehension of the method, Figure~\ref{fig:scheme_iterRSBW} presents a flowchart of the computational procedure, specifically illustrating the scenario where two consecutive RS treatments are applied, followed by a final BW resolution. 
\textcolor{black}{This computational scenario will arise  in the application of the method given in the next section.}
Additionally, note that we also provide a pseudo-code (see Algorithm \ref{code:RSBW_pseudoalgo}) that details the key numerical steps of the RSBW method. In the next section, we examine the robustness of this strategy on model Hamiltonians.}


\linespread{1.}

\begin{algorithm}[t!] 
\caption{ multi-step RSBW method }
\begin{algorithmic}[1]

\State {$\Longrightarrow$ \underline{\textbf{Step 1}}: \textit{Initialization}}

\State {{Set}} $\rho_{\rm min} \leftarrow cst$ (e.g. 0.5)

\State {Define} $\hat{H}_0$ and $\hat{W}_0$ such that $(H_0)_{ij}=\delta_{ij}H_{ij}$ and $\hat{W}_0 = \hat{H} -  \hat{H}_0$

\State Assign the eigenvectors of  $\hat{H}_0$ to $\{\ket{i}\}$ 
\State Assign  $\mathcal{E}_i \leftarrow \bra{i} \hat{H}_0  \ket{i}$ 

\State
\State {$\Longrightarrow$ \underline{\textbf{Step 2}}: \textit{Iterative RS treatment}}

\State Set $n \leftarrow  1$
\State Set $i \leftarrow  0$

\Loop { over $i$ }
  \State  $P_n$ = $\{\ket{i}\}$
  \State dim$(P_n)$ = 1
  \Loop { over $j>i$}
     \State Assess $ \rho_{ij,n}$ based on Eq.~(\ref{eq:RS_criterion_step1}) 
     \If { $\mathcal{E}_i=\mathcal{E}_j$ \textbf{or} $ \rho_{ij,n}  > \rho_{\rm min}$   } 
       \State  $P_n$ += $\{\ket{j}\}$
       \State dim$(P_n)$ += 1
     \EndIf
     \EndLoop
     \If { dim$(P_n) > 1$}  
       \State Build the orthogonal $Q_n$-space
       \State Build $\&$ diagonalize $\hat{H}_{{\rm eff},n}^{(2)}$
       \State {Build} $\hat{H}_n^{\mbox{\tiny RS}}$ and $\hat{W}_n$ following Eq.~(\ref{eq:HRS_iteration_n})
       \State Assign the eigenvectors of  $\hat{H}_n^{\mbox{\tiny RS}}$ to $\{\ket{i}\}$ 
       \State Assign  $\mathcal{E}_i \leftarrow \bra{i} \hat{H}_n^{\mbox{\tiny RS}}  \ket{i}$
       \State Set $n \leftarrow  n + 1$
      \State Set $i \leftarrow  0$
    \EndIf

\EndLoop

\State
\State {$\Longrightarrow$ \underline{\textbf{Step 3}}: \textit{ Final BW treatment}}
\State Compute all $E^{\mbox{\tiny RSBW}}_{i,n}$ following Eq.~(\ref{eq:BW_general2})
    
\end{algorithmic}
 \label{code:RSBW_pseudoalgo}
\end{algorithm}

\section{Spectroscopy from multi-step RSBW on model Hamiltonians}
The efficiency of the multi-step RSBW treatment was examined on systems ruled by parametrized Hamiltonians.
These inspections allow one to vary a selection of parameters accounting for the one-
and two-electron integrals of the electronic Hamiltonian. However, the number of parameters
was limited to set up intentionally non-trivial spectroscopies (quasi-degeneracies, presence of intruder states)
and to highlight the important contributions in the successive partitionings
of the Hamiltonian. 
The energy evaluations were compared to the exact diagonalization eigenvalues.

In practice, we start with
a  strictly degenerate model space $P_1$
spanned by two reference functions $\ket{\alpha}$ and $\ket{\alpha'}$ ($d_1 = 2$),
in the presence of two perturbers $\ket{\beta}$ and $\ket{\beta'}$ defining
the orthogonal $Q_1$-space.
Limiting the number of perturbers  does not harm the generality and 
offers  a  playground to assess the relevance of the here-proposed approach. 
The unperturbed Hamiltonian energies $\braket{\alpha|\hat{H}_0|\alpha}=\braket{\alpha'|\hat{H}_0|\alpha'}$
were set to zero, and $\braket{\beta|\hat{H}_0|\beta} = U$ and $\braket{\beta'|\hat{H}_0|\beta'} = U'$ are
the  positive energies of the perturbers (\textit{e.g.} charge transfer state energies).
Then, the perturbation was introduced by switching on 
intra and inter-space couplings.
Standard notations of molecular magnetism and Hubbard model
were used with 
$\braket{\alpha|\hat{W}_0|\alpha'}=K_\alpha$, $\braket{\beta|\hat{W}_0|\beta'}=K_\beta$,
while the number of parameters for the inter-space
couplings was limited with
$\braket{\alpha|\hat{W}_0|\beta}=\braket{\alpha|\hat{W}_0|\beta'}=t$
and $\braket{\alpha'|\hat{W}_0|\beta}=\braket{\alpha'|\hat{W}_0|\beta'}=t'$. 
Note that we will consider $t$ as the energy unit throughout the numerical inspections. 
The matrix structure of the total Hamiltonian is given by:
\begin{equation}
\renewcommand{\arraystretch}{1.5}
    \hat{H}=
    \begin{pmatrix} 
     0         &  K_\alpha  &  t       &  t      \\
     K_\alpha  &  0         &  t'      &  t'     \\
     t         &  t'        &  U       &  K_\beta \\
     t         &  t'        &  K_\beta &  U'     \\
    \end{pmatrix}.
    \label{eq:H_matrix}
\end{equation}
Let us first qualitatively analyze the structure of the Hamiltonian.
Following Bloch theory,~\cite{lindgren2012atomic} the structure of the second-order effective Hamiltonian 
acting in the $P_1$-space is given by : 
\begin{eqnarray}\renewcommand{\arraystretch}{1.5}
    \hat{H}_{\rm eff,1}^{(2)}=
     \begin{pmatrix} 
   -t^2\big (\frac{1}{U}+\frac{1}{U'} \big )  &  K_\alpha -tt'\big (\frac{1}{U}+\frac{1}{U'} \big )    \\
   K_\alpha -tt'\big (\frac{1}{U}+\frac{1}{U'} \big )  &  -t'^2\big (\frac{1}{U}+\frac{1}{U'} \big ) 
    \end{pmatrix}.
    \label{eq:Heffectif_matrix}
\end{eqnarray}
Its diagonalization  gives access to the  eigenenergies
$E^{\rm RS}_{0,1}$ and $E^{\rm RS}_{1,1}$
and their corresponding 
eigenvectors
$|\Psi^{\rm RS}_{0,1} \rangle$
and
$|\Psi^{\rm RS}_{1,1} \rangle$.
As soon as $K_\alpha \approx U$, 
one of the $\hat{H}_{\rm eff}^{(2)}$ eigenvalues
gets quasi-degenerate with the perturber
$\ket{\beta}$ (see Figure~\ref{fig:scheme_iterRSBW}, first RS treatment).  
Evidently,  such scenario questions the relevance of a BW approach to accurately hierarchize the different states
(Section~\ref{sec:SS RSBW}). Thus, a multi-step RS treatment might be desirable to guarantee 
an improved BW  expansion convergence  of each individual energy (Section~\ref{sec:MS RSBW}).
These problematic situations can be explored by varying a single parameter $K_\alpha$.
The fixed parameters defining the full Hamiltonian of the system are given in Table~\ref{table:parameters}.

 \begin{table}[H]  
  \centering
  \begin{tabular}{c|c|c|c}
    \toprule

          $U$    &    $U'$   & $K_\beta$ &   $t'$   \\  \hline \hline \toprule
          3.0    &     6.0    & 1.5      &   -1.5      \\ \hline

  \end{tabular}
  \caption{Parameters defining the full Hamiltonian ruling a system initially consisting of two model configurations and two perturbers. The model space $P_1$ configurations energies are used as a reference. All energies are given in $|t|$ unit ($t=-1$).}
  \label{table:parameters}
\end{table} 

The accuracy of the method is evaluated 
from $\Delta = {\rm sup}_{i,n} \{|E^{\rm RSBW }_{i,n} - E^{\rm exact }_{i}| \}$,  where 
 $E^{\rm exact }_i$ is the exact eigenenergy of the $i$-th eigenstate. 
The BW expansion  was fixed to $m=5$ (see Eq.~(\ref{eq:BW_general2})), limiting a $\Delta$ value to 0.1 ($t$ unit).
Finally, five iterations were sufficient to reach convergence in the self-consistent treatment {for all the $K_\alpha$ values we have considered in this study}.

\subsection{Single-step RSBW, $n=1$}
\label{sec:SS RSBW}

\begin{figure}[H] 
     \centering
     \includegraphics[width=6.5cm]{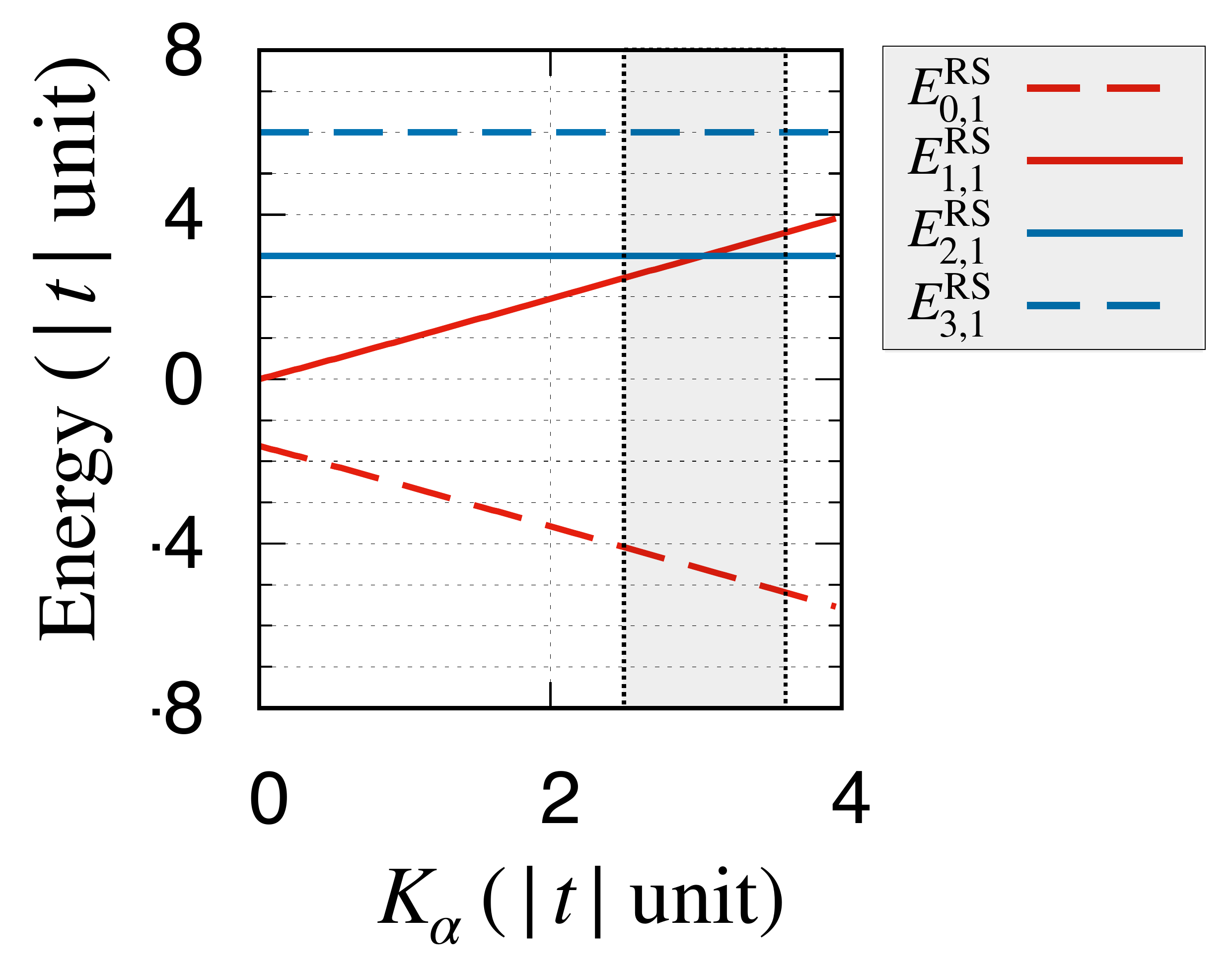}
     \caption{Variations  of the two model states energies as a function of the coupling
     parameter $K_\alpha$ after the first RS treatment for $\rho_{\rm min} \gg 1 $. The perturbers energies remain unchanged. The parameters
     values are given in Table~\ref{table:parameters}. The critical regime is highlighted.}
     \label{fig:E-RS1}
 \end{figure}

For $\rho_{\rm min} \gg 1 $, $d_1 = 2$  and a self-consistent BW treatment 
follows the diagonalization of the effective Hamiltonian built on the strictly degenerate model space $P_1$. 
The resulting eigenstates and eigenenergies $\{E^{\rm RS}_{i,1} \}_{i=0-3}$  are used as zeroth-order states in the BW expansion~\cite{delafosse2024two}. Their variations with respect to $K_\alpha$ are shown in Figure~\ref{fig:E-RS1}
as a function of the coupling parameter $K_\alpha$.

\begin{figure}[H] 
     \centering
     \includegraphics[width=7cm]{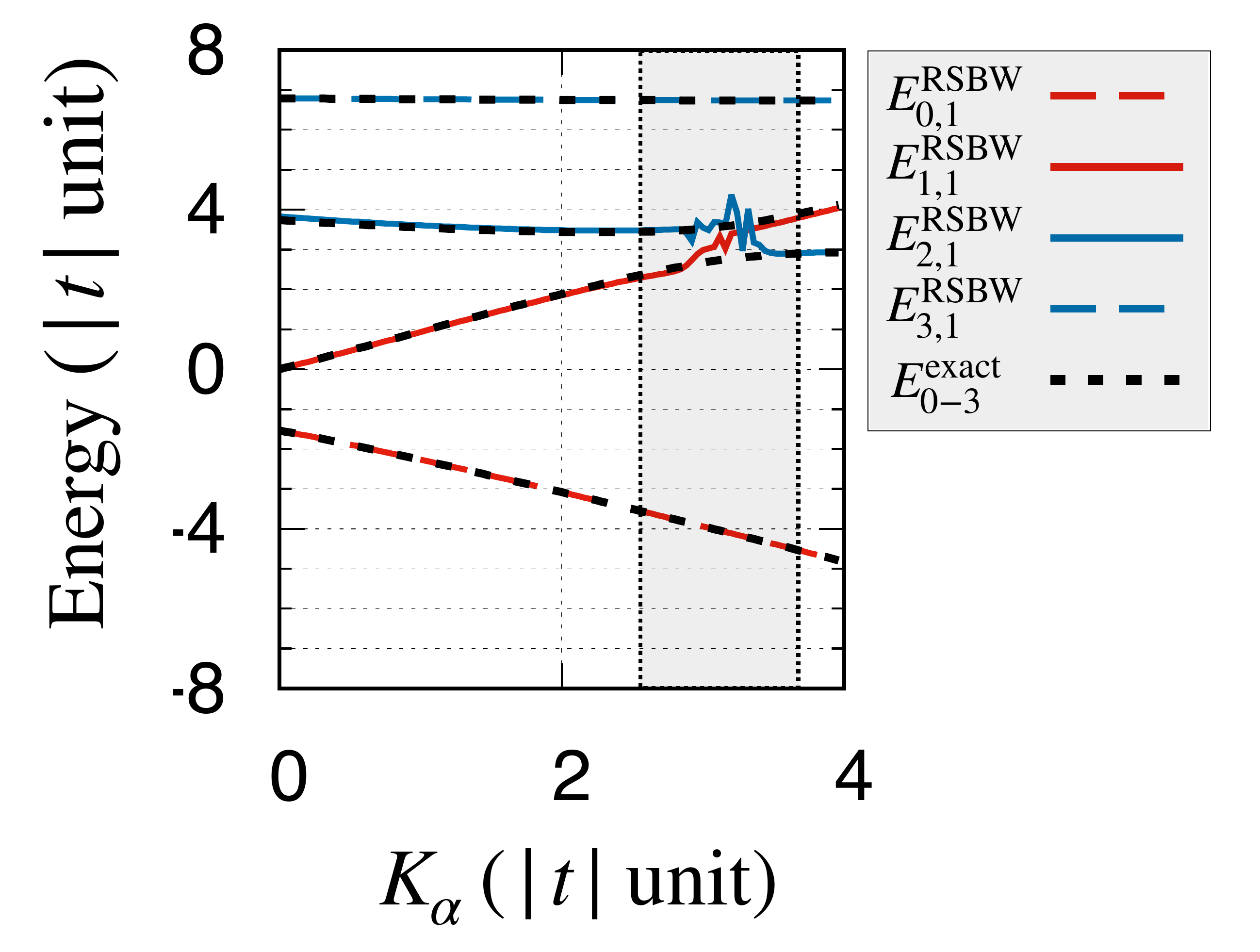} 
     \caption{Approximate eigenenergies $\{E^{\mbox{\tiny \rm RSBW}}_{i,1} \}_{i=0-3}$ as a function of $K_\alpha$ using a single-step RSBW treatment. $E^{\rm exact}_{0-3}$ refer to the exact energies obtained from the  Hamiltonian matrix exact diagonalization.}
     \label{fig:RSBW1}
 \end{figure}

This preliminary inspection suggests that in the $K_\alpha \approx U$ regime, referred to as the critical regime in the following (shaded area in Figure~\ref{fig:E-RS1}),  the state-specific BW procedure may lead  to convergence
issue due to quasi-degeneracies of the first and second excited states. 
Figure \ref{fig:RSBW1} shows the evolution of $E^{\rm RSBW}_{1,1}$ and $E^{\rm RSBW}_{2,1}$ as a function of $K_\alpha$,
featuring different behaviours.

\begin{figure}[H] 
     \centering
     \includegraphics[width=7cm]{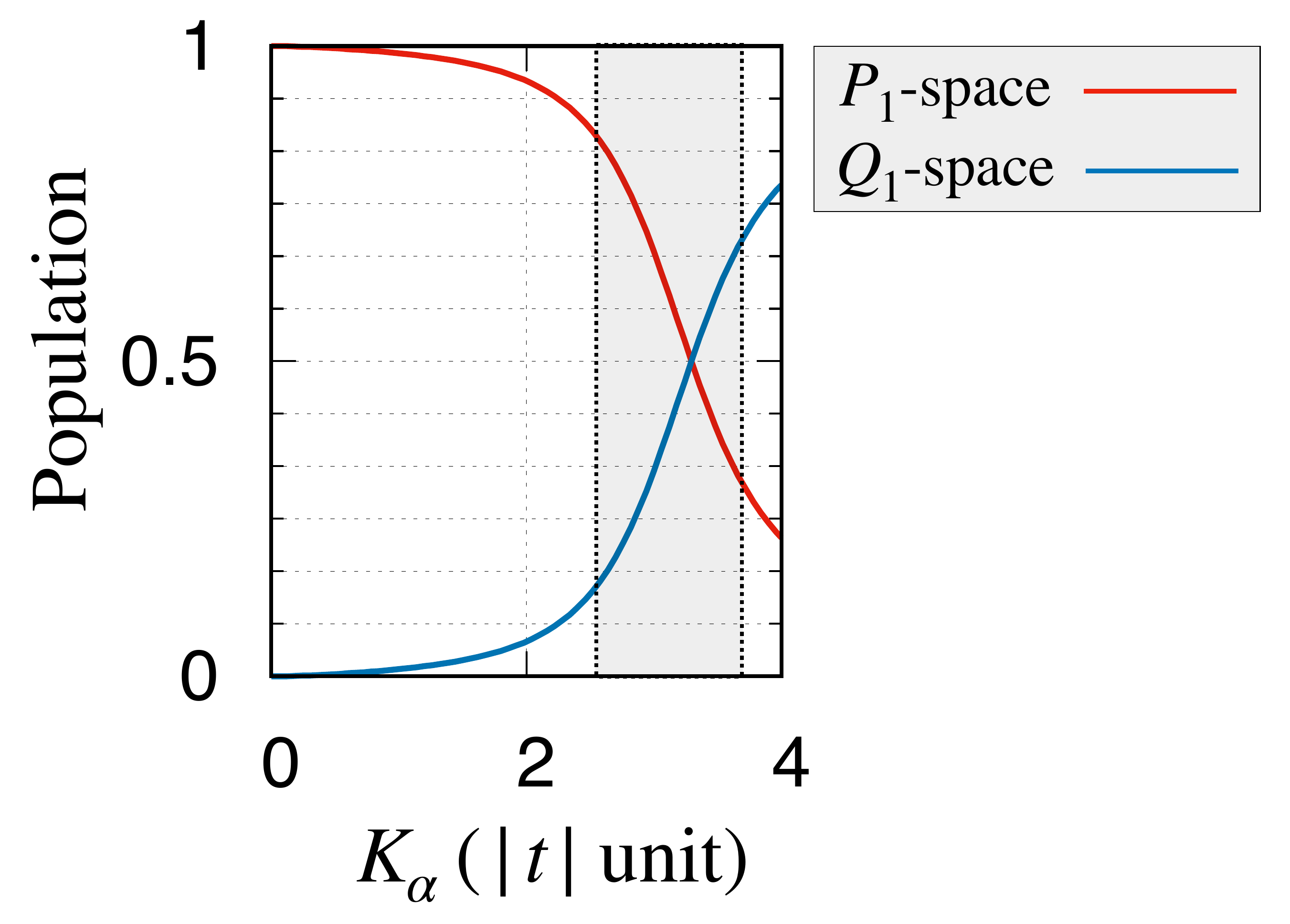}
     \caption{$P_1$- and $Q_1$-spaces populations of the first excited state obtained from the exact diagonalization of
     the $\hat{H}$ matrix given in Eq.~(\ref{eq:H_matrix}).
    }
    \label{fig:pop1-sum-a}
\end{figure}

Starting from  $K_\alpha =0$, the energies  $E^{\rm \mbox{\tiny RSBW}}_{1,1}$ and $E^{\rm \mbox{\tiny RSBW}}_{2,1}$ obtained with 
a single-step RSBW treatment are in 
very good agreement with the exact ones.
In contrast, the BW expansion breakdowns as reflected by the oscillating  energies   in the  $K_\alpha \approx U = 3$ regime
(shaded area in Figure~\ref{fig:RSBW1}).
Even for $m>5$ (see Eq.~(\ref{eq:BW_general2})), no convergence in the expansion 
is observed.
In perturbation theory, $\ket{\beta}$ is to
be considered as an intruder state 
for the model state spanned by $\ket{\alpha}$
and $\ket{\alpha'}$.
As expected, strongly coupled diabatic states
result in the well-known avoided crossing 
picture. 
Accordingly, the wavefunctions nature changes in the vicinity of this point.
Whereas the first excited state $\ket{\Psi^{\rm RS}_1}$  is initially dominated by its projection onto 
$| \alpha \rangle$ and $| \alpha' \rangle$ (\textit{i.e.} $P_1$-space),
the  $| \beta \rangle$ and $| \beta' \rangle$ (\textit{i.e.} $Q_1$-space)
predominantly contribute after the crossing (see Figure~\ref{fig:pop1-sum-a}).
Evidently, such phenomenon frequently observed in physical chemistry (mixed-valence compounds, photo-chemical processes) 
calls for particular care in electronic structure calculations.
Finally, for higher $K_\alpha$ values, agreement is recovered for the first and second
excited states energies.

Despite its performance in
improving perturbation approaches,
the RSBW method based on a single-step RS treatment is unable to correctly describe 
all the eigenenergies for identified critical regimes.
Thus, the next section explores a systematic way to reach a well-tempered BW energy expansion.

\subsection{Multi-step RSBW approach, $n>1$}
\label{sec:MS RSBW}

\begin{figure}[H] 
     \centering
     \includegraphics[width=7cm]{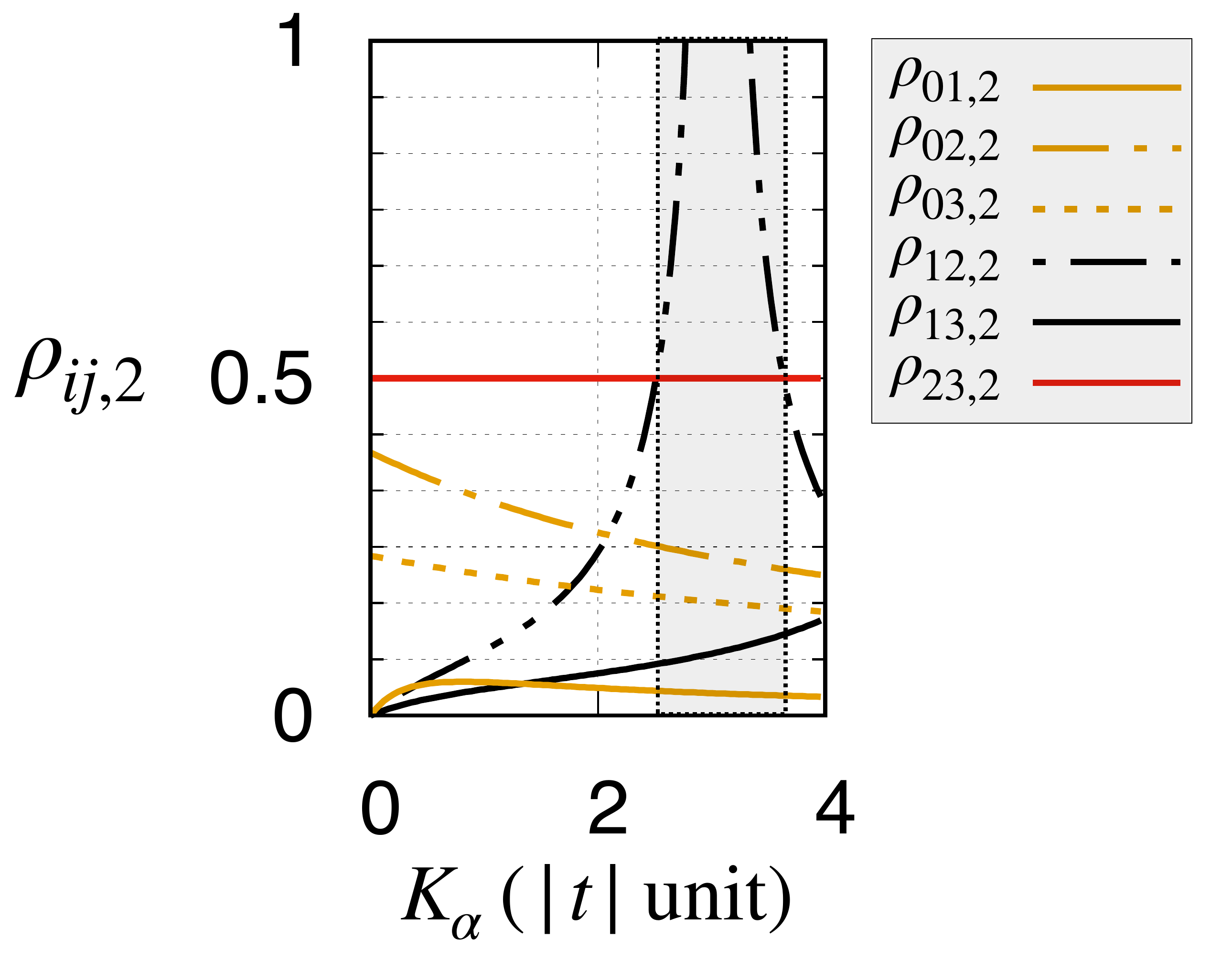}
     \caption{$\rho_{ij,2}$ values (see Eq.~(\ref{eq:RS_criterion_step_n}) for $n=1$) after the first RS step as a function of $K_\alpha$.  The shaded area shows the critical regime defined by the criterion  ${\rm sup}\{\rho_{ij,2}\} > \rho_{\rm min}$ where an arbitrary $\rho_{\rm min} = 0.5$ is used to build the $P_2$ model space. Note that  $\rho_{23,2}= \left\lvert 
    \frac{ K_\beta } {U' - U} \right\rvert$ remains constant whatever $K_\alpha$.
     }  
     \label{fig:rho-RS1}
 \end{figure}

Let us  start with the zeroth-order Hamiltonian $\hat{H}_1^{\mbox{\tiny RS}}$ 
and the perturbation $\hat{W}_1=\hat{H}-\hat{H}_1^{\mbox{\tiny RS}}$. 
The efficiency of the BW treatment  depends on the ratio $\rho_{\rm min}$ introduced in Eq.~(\ref{eq:RS_criterion_step_n}):
the larger the $\rho_{\rm min}$ value, the smaller the model space size, 
at the cost of
a slower convergence of the BW perturbative expansion.  
Thus, its value is to be controlled, possibly changed along the procedure 
to maintain tractable model
space sizes.
 Figure~\ref{fig:rho-RS1} shows the $\rho_{ij,2}$ values  calculated 
 after the first RS step as a function of  $K_\alpha$.
 
As expected from the previous discussion, a divergence  
in $\rho_{12, 2}$ is observed for $K_\alpha \approx U = 3$.
In the critical regime  defined by an arbitrary $\rho_{\rm min} = 0.5$ (shaded area in Figure~\ref{fig:rho-RS1}), a single  model space $P_2$  is identified
with $d_2 = 2$,
spanned by the $\ket{\Psi^{\mbox{\tiny RS}}_{1,1}}$ and $\ket{\Psi^{\mbox{\tiny RS}}_{2,1}}$
states.

Thus, a second RS treatment must be  performed to account for this
strong mixing marked by the $\rho_{12,2}$ value. 
Since the zeroth-order energies 
$E^{\mbox{\tiny RS}}_{1,1} = \braket{\Psi^{\mbox{\tiny RS}}_{1,1}|\hat{H}^{\mbox{\tiny RS}}_{1}|\Psi^{\mbox{\tiny RS}}_{1,1}}$
and
$E^{\mbox{\tiny RS}}_{2,1} = \braket{\Psi^{\mbox{\tiny RS}}_{2,1}|\hat{H}^{\mbox{\tiny RS}}_{1}|\Psi^{\mbox{\tiny RS}}_{2,1}}$
may differ,
the updated second-order effective Hamiltonian $\hat{H}_{\rm eff,2}^{(2)}$   might not be
hermitian. To remedy this possible difficulty, a common value  
$(E^{\mbox{\tiny RS}}_{1,1}+E^{\mbox{\tiny RS}}_{2,1})/2$ was used and the perturbation $\hat{W}_2$
was defined accordingly $\hat{W}_2 = \hat{H} - \hat{H}^{\mbox{\tiny RS}}_{2}$ (see Eq.~(\ref{eq:HRS_def})).
At this stage, let us stress that the 
$Q_2$-space is spanned by the ground and third excited states,  $\ket{\Psi^{\mbox{\tiny RS}}_{0,1}}$ and
$\ket{\Psi^{\mbox{\tiny RS}}_{3,1}}$ (see Figure~\ref{fig:scheme_iterRSBW}).
It is a rather unusual picture since all perturbers traditionally lie 
higher in energy.



For $\rho_{12,2} > \rho_{\rm min}$, the diagonalization of the effective Hamiltonian $\hat{H}_{\rm eff,2}^{(2)}$  
defined in the $P_2$-space gives access to the new zeroth-order eigenenergies $E^{\mbox{\tiny RS}}_{1,2}$ and $E^{\mbox{\tiny RS}}_{2,2}$ and their corresponding eigenvectors 
$\ket{\Psi^{\mbox{\tiny RS}}_{1,2}}$  and  $\ket{\Psi^{\mbox{\tiny RS}}_{2,2}}$.
From these quantities and $\hat{W}_2$, one can easily evaluate the $\{\rho_{ij,3}\}$ values.
Interestingly, by concentrating part of the perturbation in the successive
redefinition of the zeroth-order Hamiltonians
$\hat{H}_1^{\mbox{\tiny RS}}$ and $\hat{H}_2^{\mbox{\tiny RS}}$,
 the ratio $\rho_{12,3}$ takes finally lower values than
$\rho_{12,2}$, and even lower than $\rho_{\rm min} = 0.5$. 
Since a similar conclusion holds for all $\{\rho_{ij,3}\}$,
no further RS treatment is required at this stage
of the procedure.

\begin{figure}[!ht] 
     \centering
     \includegraphics[width=6.8cm]{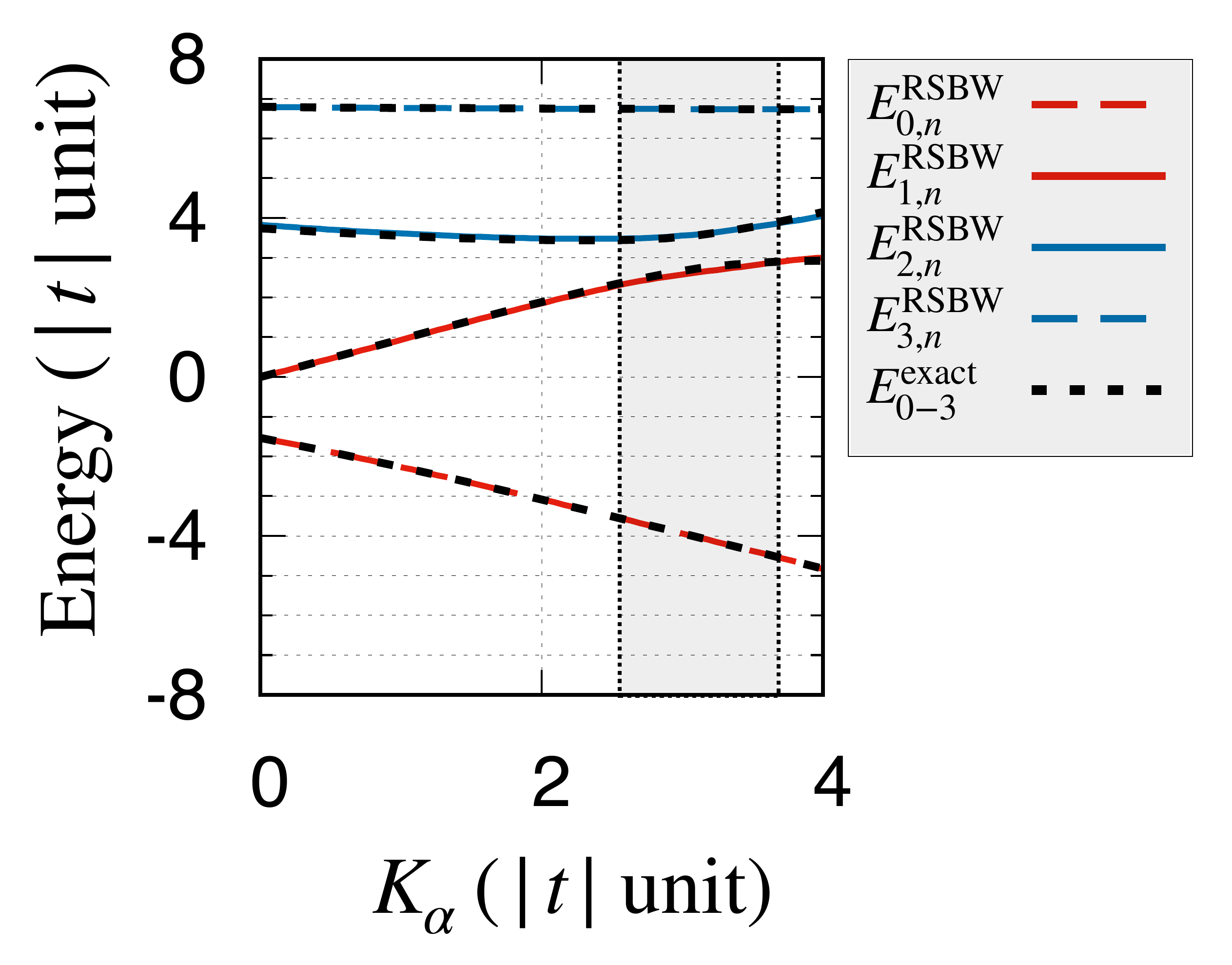}     

     \caption{Approximate eigenenergies $\{E^{\mbox{\tiny \rm RSBW}}_{i,n} \}_{i=0-3}$ as a function of $K_\alpha$ using a single- ($n$=1) or two-step ($n$=2, shaded area) RSBW treatment. $E^{\rm exact}_{0-3}$ refer to the exact energies obtained from the  Hamiltonian  matrix exact diagonalization.}
     
     \label{fig:E-RS2-BW}

\end{figure}

As mentioned before, a BW expansion limited to fifth-order  is conducted as the last step of the calculation. 
As seen in Figure~\ref{fig:E-RS2-BW}, agreement with
exact energies ($\Delta = 0.097$)
is observed in the whole range of
$K_\alpha$ values.
Let us stress that for the ground and third-excited states, this accuracy is already reached with a single-step RS
and a BW expansion limited 
to third-order ($m=3$).
In contrast, the first and second excited states call
for a multi-step RSBW approach  to overcome the intruder
state issue (shaded area in Figure~\ref{fig:E-RS2-BW}) 
and a $m=5$ expansion to reach the desired accuracy (i.e. $\Delta < 0.1$).


The failure of the single-step RS procedure triggered
by the $K_\alpha$ parameter is
immediately identified by the divergence of  
 $\rho_{12,1}$ (see Figure~\ref{fig:rho-RS1}).
Evidently, such scenario can be encountered for other sets
of parameters ruling the Hamiltonian.
However, the most important concern in the procedure is  to check  the
partitioning quality $\hat{H} = \hat{H}^{\mbox{\tiny RS}}_n + \hat{W}_n$ after each RS step
through the evaluations of the
$ \{\rho_{ij,n+1}\}$ ratios (see Eq.~(\ref{eq:RS_criterion_step_n})).
Whatever the system, the screening of interacting states
can be carried out successively.
As soon as ${\rm sup}\{\rho_{ij,n+1}\} < \rho_{\rm min}$, the subsequent BW  treatment is
improved,
and a better description of the system's eigenenergies is garanteed at a lower numerical cost.
Along the procedure, one may at will vary the $\rho_{\rm min}$ threshold to define a new critical regime and build up the
associated model spaces $P_n$. 
This particular
flexibility is left for future inspections.

\section{Conclusion}
The multi-step RSBW method we proposed  combines successive
Rayleigh-Schrödinger (RS) treatments leading to a well-tempered 
state-specific Brillouin-Wigner 
(BW) expansion. 
The strategy is  to progressively include the perturbation effects
by building successively second-order effective Hamiltonians.
Starting from the low-lying states, a first effective Hamiltonian is built using a
first-order perturbation criterion. The resulting model functions
and eigenenergies include contributions of the perturbation,
and the definition of the zeroth-order Hamiltonian  is revisited.
Based on this updated partitioning of the Hamiltonian, the procedure is
then repeated to carefully account for strongly interacting
states and the presence of quasi-degeneracies. 

The magnifying-glass scanning performed on sub-spaces not only  allows
one to concentrate the effort on some particular energy windows,
but also to progressively reduce the impact of the resulting
perturbation and the size-consistency error. 
Finally,  a systematic and numerically cheap order-by-order BW expansion
is performed on each individual state (state-specific) and leads to accurate energy
transitions.
The relevance of the multi-step RSBW strategy is highlighted on model Hamiltonians.
The sizes of the model Hamiltonians are controlled by
a threshold parameter which is to be further explored by
implementing the here-proposed method on the quantum chemistry Hamiltonian.

\section{Acknowledgments}
This work was supported by the ANR (PRC ANR-2023-CE07-0035 AtropoPhotoCat).
The authors wish to thank Prof. J. Browaeys for helpful discussions regarding accuracy assessment.
 
\section*{References}
\bibliographystyle{iopart-num} 
\bibliography{biblio}

\end{document}